\documentclass[lettersize,journal]{IEEEtran}
\usepackage{amsmath,amsfonts}
\usepackage{array}
\usepackage{textcomp}
\usepackage{stfloats}
\usepackage{url}
\usepackage{verbatim}
\usepackage{graphicx}
\usepackage{cite}
\usepackage{amssymb}
\usepackage{hyperref}
\usepackage{url}
\usepackage{color}
\usepackage[T1]{fontenc}
\usepackage{booktabs} 
\usepackage{multirow} 
\usepackage{subcaption}
\usepackage{float}
\usepackage{algpseudocode}
\usepackage{algorithm}
\usepackage{threeparttable}
\usepackage{circledsteps}

\begin{document}

\title{Pk-IOTA: Blockchain empowered Programmable Data Plane to secure OPC UA communications in Industry 4.0}

\author{\IEEEauthorblockN{Lorenzo Rinieri\IEEEauthorrefmark{1}\IEEEauthorrefmark{2},
Giacomo Gori\IEEEauthorrefmark{1}\IEEEauthorrefmark{2},
Andrea Melis\IEEEauthorrefmark{1}, 
Roberto Girau\IEEEauthorrefmark{1}, 
Marco Prandini\IEEEauthorrefmark{1}, and
Franco Callegati\IEEEauthorrefmark{1}}

\IEEEauthorblockA{\IEEEauthorrefmark{1}Department of Computer Science and Engineering (DISI), University of Bologna, Bologna, Italy}

\IEEEauthorblockA{\IEEEauthorrefmark{2}These authors contributed equally to this work as co-first authors.}
}




\maketitle

\begin{abstract}
The OPC UA protocol is becoming the de facto standard for Industry 4.0 machine-to-machine communication. It stands out as one of the few industrial protocols that provide robust security features designed to prevent attackers from manipulating and damaging critical infrastructures. However, prior works showed that significant challenges still exists to set up secure OPC UA deployments in practice, mainly caused by the complexity of certificate management in industrial scenarios and the inconsistent implementation of security features across industrial OPC UA devices.
In this paper, we present Pk-IOTA, an automated solution designed to secure OPC UA communications by integrating programmable data plane switches for in-network certificate validation and leveraging the IOTA Tangle for decentralized certificate distribution. Our evaluation is performed on a physical testbed representing a real-world industrial scenario and shows that Pk-IOTA introduces a minimal overhead while providing a scalable and tamper-proof mechanism for OPC UA certificate management.
\end{abstract}


\section{Introduction}
\label{sec:introduction}
In the landscape of Industrial Control Systems (ICS), which integrate cyber and physical components to control the operations of critical infrastructures, it is paramount to ensure secure and reliable communication between devices~\cite{ike2023scaphy}. Furthermore, with the advent of the Industry 4.0 and the Industrial Internet of Things (IIoT), machines, devices, sensors and actuators are becoming increasingly interconnected and capable of
communication with each other. The OPC Unified Architecture (OPC UA) protocol~\cite{OPC10000-1} has emerged as the de-facto standard for facilitating interoperability and data exchange between devices from different manufacturers~\cite{kohnhauser2021security}.
OPC UA was designed with built-in security mechanisms that protect the authenticity, integrity, and confidentiality of data in transit and is attested secure by the German Federal Office for Information Security~\cite{BSI_Sec_An}.

Despite the sophisticated security features embedded within the OPC UA standard, recent studies highlight a significant security gap between OPC UA's design and its real-world deployments due to its configuration complexity. Dahlmanns et al.~\cite{dahlmanns2020easing} revealed that 92\% of OPC UA deployments accessible over the Internet are improperly configured, due to missing access control, disabled security functionality, use of deprecated cryptographic primitives, or certificate reuse. Similarly, Erba et al.~\cite{erba2022security} systematically assessed 48 real-world OPC UA artifacts, uncovering widespread deficiencies in security features, certificate management, and trust list handling, which expose OPC UA systems to various attacks such as feeding incorrect information to clients or eavesdropping and changing values which can directly alter the physical process.

These findings underscore two primary challenges in practical OPC UA deployments: the complexity of certificate and trust list management and the inconsistent implementation of security features across industrial devices. To address them, we present Pk-IOTA, an automated architecture that integrates the programmable data plane with a decentralized PKI based on the IOTA Tangle to secure OPC UA deployments. Pk-IOTA enforces real-time certificate validation at the network level, thereby offloading security responsibilities from individual OPC UA devices. Additionally, by utilizing the immutable and tamper-proof nature of blockchain, Pk-IOTA eliminates the manual overhead associated with certificate handling, ensuring real-time trust verification, reducing the potential for misconfiguration, and offering a scalable and resilient solution that is compatible with a broad spectrum of OPC UA devices, even those with limited native support for advanced security features.
We evaluate Pk-IOTA using a physical testbed that simulates real-world Industry 4.0 OPC UA deployment in different locations, assessing the in-network overhead introduced by certificate validation, the efficiency of certificate propagation through the IOTA Tangle, and the overall resource consumption of the architecture's components. Our evaluation shows that Pk-IOTA introduces a minimal overhead of 14\% on the OPC UA security handshake without interfering with normal operations while providing a scalable and tamper-proof mechanism for certificate issuance and revocation. Additionally, resource consumption analysis demonstrates that all components of the Pk-IOTA architecture operate within feasible limits, ensuring that the system can be deployed in large-scale industrial settings without imposing significant performance penalties. Finally, we discuss how Pk-IOTA prevents the attacks presented in~\cite{erba2022security}.

In summary, our work makes the following contributions:
\begin{itemize}
    \item We propose Pk-IOTA, an architecture that leverages the programmable data plane and Distributed Ledger Technology to automate certificate management and validation in OPC UA deployments.
    \item We conduct a comprehensive evaluation of Pk-IOTA employing a physical testbed, showing its minimal impact over existing OPC UA deployments.
    \item We publicly release the implementation of Pk-IOTA\footnote{\url{https://github.com/UniboSecurityResearch/Pk-IOTA}}.
\end{itemize}

This paper is organized as follows. Section~\ref{sec:background} gives an overview of the OPC UA protocol and its security handshake. Section~\ref{sec:problem} motivates the need for Pk-IOTA while Section~\ref{sec:chall} presents its design goals and the key technological enablers. Then, Section~\ref{sec:threatmodel} outlines our assumptions about the System and Threat Model. Section~\ref{sec:pk-iota} presents an in-depth overview of the proposed Pk-IOTA architecture and Section~\ref{sec:eval} describes the experimental setup and evaluates our solution. A discussion of the Pk-IOTA architecture in terms of security, scalability, and cost is presented in Section~\ref{sec:discussion}. Finally, Section~\ref{sec:relatedwork} analyzes related work, and Section~\ref{sec:conclusion} draws our conclusions.

\section{OPC UA Background}
\label{sec:background}
OPC UA~\cite{Mahnke_Leitner_Damm:2009} is a machine-to-machine communication standard designed to ensure secure, reliable, and platform-independent exchange of data between devices and systems in Industry 4.0 settings. 
The main two features of OPC UA are its information model-based e3wegrqqarchitecture~\cite{OPC10000-5}, which enables platform-independent communication between industrial devices of different manufacturers, as well as integrated secure communication by design~\cite{BSI_Sec_An}.

\begin{figure}[ht]
    \centering
    \includegraphics[width=1.0\linewidth]{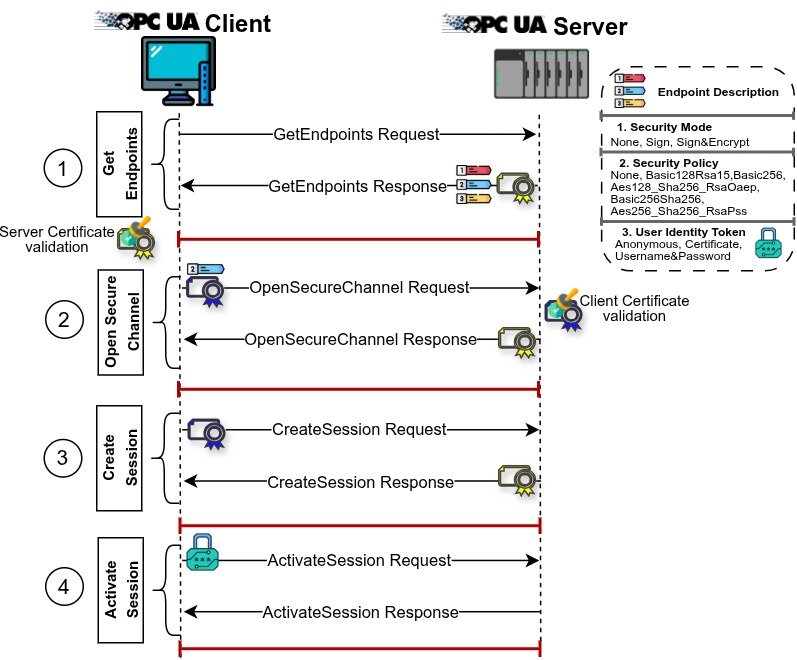}
    \caption{Connection establishment handshake in OPC UA.
    }
    \label{fig:handshake}
\end{figure}

To establish a communication channel, an OPC UA client and an OPC UA server first need to perform a security handshake made of four steps, as shown in Figure~\ref{fig:handshake}. 
In the first step, the client sends a GetEndpoints request to discover the server's available Session Endpoints, each defined by a Security Mode, a Security Policy, and the supported User Identity Token(s). The Security Mode defines how messages are exchanged between parties to achieve authentication, confidentiality, and integrity. These modes offer unprotected communication (None), authenticated communication (Sign), and authenticated as well as confidential communication (SignAndEncrypt). Security Policies define the cryptographic primitives and their parameters to realize the different Security Modes. Finally, the UserIdentityToken defines the supported user authentication methods for an endpoint: Anonymous (no user authentication), Username\&Password, and Certificate.

Upon receiving the GetEndpoints response, the client selects a Session Endpoint and validates the server's Application Instance Certificate. If the certificate is deemed trustworthy, the client sends an OpenSecureChannel request. The server then validates the client's certificate in turn and, if trustworthy, responds with the OpenSecureChannel response. The Secure Channel has a finite lifetime, after which it must be renewed without affecting the ongoing Session.

In the third step, the client creates a Session on top of this Secure Channel. Each Session is uniquely identified by a sessionId (used for logging in the server's Address Space) and an authenticationToken (used to link requests to the Session).

In the fourth step, the client activates the Session by invoking the ActivateSession service. From this point onwards, the Session remains active and all subsequent requests and responses use the symmetric keys derived in the OpenSecureChannel phase, adhering to the selected Security Mode and Security Policy.

While OPC UA's protocol design is inherently secure~\cite{BSI_Sec_An}, the range of configuration options available can greatly affect its overall security. Official guidelines~\cite{OPC_Sec_Advise} recommend always employing SignAndEncrypt Security Mode (to ensure that messages are both signed and encrypted) and disallowing anonymous authentication. Finally, only three (\textit{Aes128\_Sha256\_RsaOaep, Basic256Sha256, Aes256\_Sha256\_RsaPss}) of the six available security policies should be employed, as one offers no security (\textit{None}) and two have been deprecated due to reliance on SHA-1 (\textit{Basic128Rsa15, Basic256}).

\section{Problem Statement}
\label{sec:problem}
The majority of the security goals of OPC UA are achieved by relying on the Secure Channel~\cite{meier2022secure}, which ensures the confidentiality and integrity of exchanged data by encrypting and signing messages.
Furthermore, when establishing secure communication, OPC UA clients and servers mutually authenticate to each other by exchanging their Application Instance Certificates and generating a session key for subsequent secure communication. 
Similar to many other network protocols, OPC UA uses X.509~\cite{rfc5280} compliant certificates for authentication. In OPC UA, it is possible to use either certificates issued by a trusted Certification Authority or self-signed certificates. 
To verify received certificates, each client and server maintains a set of trustworthy certificates in its so-called trust list. A server accepts connections from a client in case the client can authenticate itself with a certificate that the server can successfully verify based on its local trust list, and vice versa~\cite{kohnhauser2022secure}.
A common approach for managing these certificates involves manually installing them into the trust lists of applications and devices. 
This method, however, is not feasible for large OPC UA networks~\cite{karthikeyan2018pki}, requires significant manual effort, and is thus prone to human error. 
Conveniently, OPC UA defines methods for the automated management of certificates~\cite{OPC10000-12}, which includes functionalities designed for standard OPC UA applications (both servers and clients), as well as capabilities tailored for a specialized server responsible for certificate management, referred to as the Global Discovery Server (GDS). The available methods enable applications to obtain certificates and update their trust lists, either through a process performed by the application itself (Pull Model) or on the initiative of the GDS (Push Model).

However, recent works revealed that significant challenges to set up secure OPC UA deployments in practice still exist. In~\cite{dahlmanns2020easing}, the authors show that 92\% of OPC UA deployments reachable over the Internet are improperly configured. In particular, 24\% of these servers completely disable communication security, while another 25\% rely on outdated cryptographic methods, such as SHA-1. Additionally, 35\% of the systems improperly implement otherwise secure configurations by reusing security-critical certificates across devices in several systems within different networks, making these susceptible to impersonation and eavesdropping. Finally, 44\% servers allow unauthenticated users to read, write, and execute functions on industrial devices. The authors affirm that secure protocols are no guarantee for secure deployments: they underscore the need to reduce configuration complexity in OPC UA deployments and demand secure defaults for all configuration options, eventually transitioning from security by design to security by default.
Diving deeper, Erba et al.~\cite{erba2022security} systematically investigated the security of 48 real-word OPC UA artifacts 
showing that many of them lack support for the security features of the protocol. Specifically, only 10 artifacts (20.8\% out of 48) correctly implement the certificate management, while 7 artifacts do not support security features at all, and the remaining 31 (64.6\% of the total) show issues or errors in the trust list management. Moreover, the features the GDS offers to manage certificates are supported by only 5 artifacts. The authors show that this inconsistent or nonexistent support for the management of trust lists and integration with the GDS can lead to misconfigurations that render OPC UA deployments vulnerable to Rogue Server, Rogue Client, and Middleperson attacks. 
Therefore, while the security-by-design of the OPC UA standard is certified by the German Federal Office for Information Security (BSI)~\cite{BSI_Sec_An}, it is evident that practical deployments of OPC UA often fail to meet these security standards due to misconfigurations, lack of support for key security features, and the complexities associated with managing certificates and trust lists in real-world ICS environments. This security gap between protocol design and practical deployments highlights the need for automated configuration processes to ensure that OPC UA systems can consistently achieve the intended level of security in operational settings.

\section{Design Goals \& Challenges}
\label{sec:chall}
To close the security gaps outlined in the previous Section, this work aims to facilitate the secure deployment and configuration of OPC UA systems in practical ICS settings. To achieve this goal, we tackle the following key challenges:
\begin{itemize}
    \item The complexity of OPC UA certificate and trust list management in real-world deployments: the manual management of certificates and trust lists in large-scale OPC UA networks is impractical and highly error-prone. Although current OPC UA setups often rely on GDS for certificate management, many ICS environments still lack GDS support, and many OPC UA industrial devices lack compatibility with GDS features~\cite{erba2022security}. At the moment of the writing, existing commercial GDS solutions are also limited in number, and none are currently certified by the OPC Foundation\footnote{\url{https://opcfoundation.org/products/?category=18}}.
    \item Inconsistent support for OPC UA security features: OPC UA provides extensive security features, but these are inconsistently implemented across devices, with many lacking critical support for certificate management, trust list handling, or GDS integration. Such inconsistency hinders the security of OPC UA deployments, as administrators cannot rely on uniform functionality across devices. Standardizing security configurations becomes challenging when devices exhibit varied support for fundamental security features.
\end{itemize}

To address these specific challenges, we present a scalable architecture with the following design goals:
\begin{itemize}
    \item Automated certificate management: by leveraging a Distributed Ledger Technology (DLT) as a decentralized PKI, we eliminate manual, error-prone certificate handling and ensure that trust relationships between OPC UA devices are accurately maintained in real-time.
    \item Independence from GDS: while GDS provides valuable certificate management functions, not all OPC UA environments support it. Therefore, our approach removes reliance on GDS by distributing OPC UA certificates over a DLT and validating certificates in the network data plane via programmable switches.
    \item Independence from device-level security feature support: by integrating in-network certificate validation via Data Plane Programmable (DPP) switches, our architecture performs real-time certificate verification that operates independently of device-specific security features such as trust list management. This design ensures a uniform security layer even if individual OPC UA devices have limited security features. We rely solely on a common set of features supported by most OPC UA devices (85.4\% out of the total number of artifacts analyzed by~\cite{erba2022security}). 
    \item Security requirements: the system is designed to be robust to Rogue Server, Rogue Client, and Middleperson attacks by mandating certificate validation at the data plane, thus blocking unauthorized devices before they access critical ICS components.
\end{itemize}

\subsection{Data Plane Programmability (P4)} 
One of the key enablers of our approach is the adoption of P4~\cite{bosshart2014p4}, an open-source programming language that enables end users to define how network packets are parsed and processed. 
P4 programs establish a pipeline consisting of parsers, match-action tables, and control flows, offering remarkable reconfigurability by allowing the dynamic redefinition of packet parsing and classification logic through the controller without necessitating changes to the physical hardware. Additionally, P4 provides protocol independence, enabling switches to handle diverse packet formats by defining custom parsers that extract header fields with user-defined names and types, alongside versatile match-action tables to process these headers. This flexibility facilitates the inspection of OPC UA messages and the extraction of certificate data directly within the switch's data plane. Furthermore, P4's target independence ensures that the same program can be compiled across various hardware or software platforms, enhancing the portability of our solution across different network infrastructures.

By deploying P4-based programmable switches, we move OPC UA certificate validation directly into the network path. Rather than relying on each endpoint device to maintain an up-to-date trust list, the network itself inspects connection requests and extracts certificates. Any OPC UA connection employing an untrusted or revoked certificate is dropped immediately, effectively preventing unauthorized connections.

\subsection{Decentralized certificate manager}
In this chapter, we motivate the use of a distributed certificate manager, analyzing possible approaches for its implementation. 

The OPC UA standard adopts a centralized certificate authority for certificate management. However, this approach introduces significant challenges, as it typically involves extensive human intervention, requiring administrators to handle certificate requests, revocations, and renewals ~\cite{patsonakis2018towards}.

A promising alternative to the centralized approach is the use of DLT, which inherently supports decentralized architectures. Blockchain, a specific version of DLTs, offers several advantages that make it a potential candidate for distributed certificate management in industrial contexts~\cite{singla2018blockchain}. Each node holds a complete copy of the blockchain, facilitating automatic synchronization across multiple controllers and clients for automated certificate management \cite{kubilay2019certledger}. 

However, blockchain faces key challenges, such as scalability and performance issues~\cite{yakubov2018blockchain}. 

For this reason, innovative solutions have been proposed to overcome the limitations of blockchain, mainly elaborating different consensus protocols or validation mechanisms inside the P2P network~\cite{yadav2023evolution} and changing the chain-like architecture: a promising project in this direction is IOTA.\\
"The Tangle"~\cite{tangle2018} introduced the mathematical foundations of the IOTA project. It focused on providing a distributed microtransaction infrastructure for the IoT domain, designed to be more energy-efficient and faster than classical blockchains, by leveraging a Direct Acyclic Graph (DAG).
In IOTA, transactions are validated through mutual confirmation, without the need for block formation, improving the speed of the transaction by ensuring that at least partial peer confirmations occur quickly~\cite{yeow2017decentralized}, thus avoiding the bottleneck that is caused by many blockchain architectures and enabling asynchronous and parallel transactions~\cite{fan2021performance}.

Although IOTA has an innovative DAG-based architecture, it preserves almost all of the benefits of traditional blockchains, such as protection from tampering, immutability, the absence of a central point of failure, and transparency \cite{zheng2018blockchain}, while also introducing enhanced scalability and efficiency: key improvements that were essential for addressing the limitations of traditional blockchain systems \cite{conti2022survey}.
These characteristics let us designate IOTA's technologies as a secure, efficient, and tamper-proof choice to implement a distributed certificate manager \cite{wang2022dag}.

\section{System \& Threat Model}
\label{sec:threatmodel}
\begin{figure}
    \centering
    \includegraphics[width=0.85\linewidth]{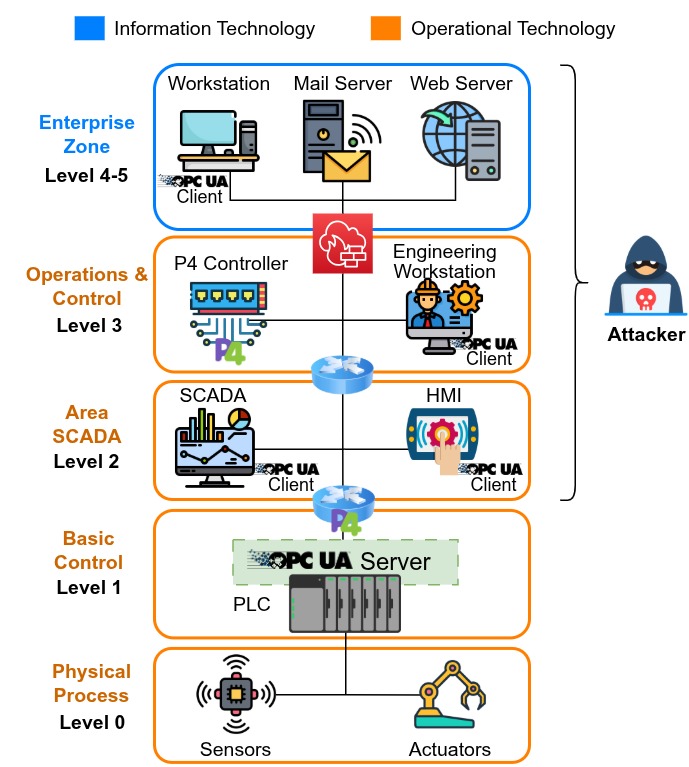}
    \caption{The Threat Model of this work, based on the Purdue Reference Architecture.}
    \label{fig:threatmodel}
\end{figure}
We illustrate the system and threat model behind this work through the Purdue Enterprise Reference Architecture \cite{williams1994purdue} for ICS systems security. As depicted in Figure~\ref{fig:threatmodel}, the Purdue Architecture organizes the ICS network into six layers: levels 4 and 5 form the Information Technology (IT) network, while the lower levels constitute the Operational Technology (OT) network. The latter handles the control, monitoring, and automation of physical processes. 
At level 0, sensors and actuators are deployed to interact with the physical process: we refer to them as IIoT devices. They are directly connected to level 1, which comprises various Programmable Logic Controllers (PLCs). The PLCs implement systems control logic by observing sensor readings and consequently updating actuator signals. We assume they expose an OPC UA server and employ the OPC UA protocol for communicating with the upper layers. This aspect has become so ubiquitous in the Industry 4.0 ecosystem that virtually every major PLC vendor today includes an embedded OPC UA in their products~\cite{arc2018opc, industryarc2022opc, industryarc2024opcua}.
Levels 2 (SCADA, HMI) and 3 encompass devices that implement supervisory control, data acquisition, and monitoring to manage plant operations. To provide these functionalities, these devices act as OPC UA clients by requesting sensors and actuators data collected by the OPC UA server at level 1. As described in Section~\ref{sec:background}, we suppose that OPC UA communication is properly configured as described by the official OPC Foundation guidelines~\cite{OPC_Sec_Advise}, i.e. (1) by enabling Security Mode SignAndEncrypt with certificates either signed by a trusted Certification Authority or self-signed certificates, (2) disabling anonymous authentication, and (3) employing one of the three suggested Security Policies.
In our system model, we finally assume that network communication between levels 1 and 2 is enabled through a P4 Programmable Switch. The P4 controller is deployed at level 3 and is able to communicate with the Enterprise Zone and with IOTA through the demilitarized zone (DMZ), which manages the connection between the IT and the OT networks while keeping them isolated from each other~\cite{iso62443-3-3}.

With respect to the attacker model, we assume that the ICS network can be partially controlled by a Dolev--Yao intruder~\cite{dolev1983security}. A typical Dolev--Yao intruder can access all public network messages and modify, inject, delete, or delay them. The Dolev-Yao attacker has, therefore, almost unlimited capabilities. However, the intruder is constrained by the perfect cryptography assumption: he can only decrypt a ciphertext or forge a signature if they possess the corresponding keys. Therefore, cryptographic attacks (e.g. brute force on Private Keys or dictionary attacks on passwords) cannot be carried out by the attacker. Additionally, aligning with previous works in ICS security~\cite{abbas2024sain,ike2023scaphy, pickren2024compromising,erba2022assessing}, we assume the attacker is able to operate only above level 1 of the Purdue Enterprise Reference Architecture. The last assumption that limits the Dolev-Yao attacker is that we assume the channel between the P4 controller and the P4 programmable switch to be secure~\cite{yigit2019secured}; thus, the intruder cannot tamper the P4 Runtime Southbound APIs in any way.
The attacker's goal is to perform the following attacks. (1) Rogue Server: a new OPC UA device has been added to the network and needs to establish secure OPC UA communication with other devices. The attacker on the network aims to deceive this new OPC UA client by injecting malicious data. The attacker sets up an OPC UA server offering secure endpoints, making new clients believe they are communicating with the legitimate OPC UA server. (2) Rogue Client: the attacker set up a client that, despite lacking authorization from the network administrator, tries to establish a connection with the server to eavesdrop or alter the information exchanged. (3) The Middleperson attack requires achieving both Rogue Client and Rogue Server objectives.
Our attacker model is consistent with those in the prior work of OPC UA security~\cite{erba2022security,puys2016formal,dreier2017formally,dreier2019formally,kohnhauser2021security}.

\section{Pk-IOTA}
\label{sec:pk-iota}
We introduce Pk-IOTA, an automated architecture designed to manage and distribute OPC UA certificates, addressing the challenges discussed in Section~\ref{sec:chall}. We first provide a comprehensive overview of the elements encompassed within our architecture, along with the underlying assumptions, and then we outline the workflow illustrating how entities interact during runtime.

\subsection{Architecture}
\label{subsec:architecture}
\begin{figure}
    \centering
    \includegraphics[width=0.9\linewidth]{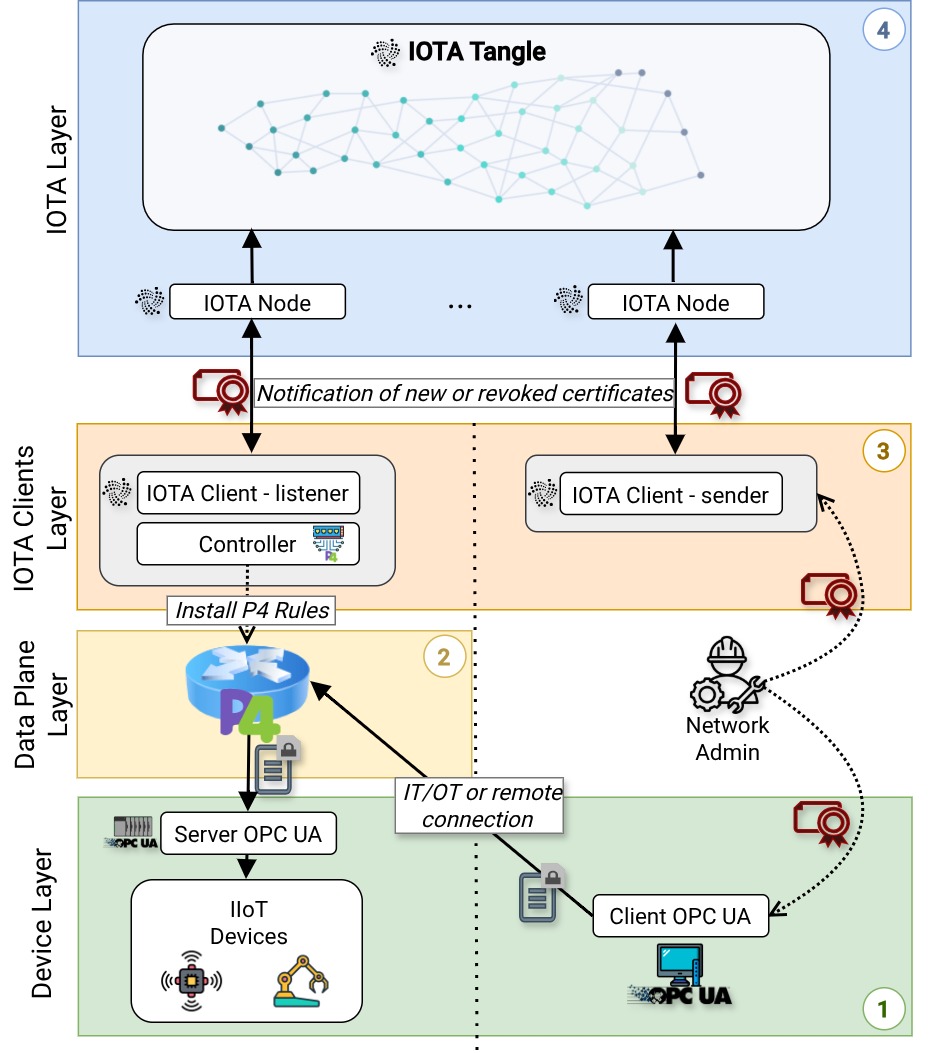}
    \caption{Overview and interactions between the 4 layers of the Pk-IOTA architecture.}
    \label{fig:arch_pkiota}
\end{figure}

The architecture of Pk-IOTA, represented in Figure~\ref{fig:arch_pkiota}, is composed of four key and interconnected layers: \Circled{1} Device Layer, \Circled{2} Data Plane Layer, \Circled{3} IOTA Clients Layer, and \Circled{4} IOTA Layer. Its foundation is the Device Layer, comprising OPC UA clients and servers deployed in accordance with the Purdue Enterprise Reference Architecture. Building on this, the Data Plane Layer integrates P4 programmable switches to enforce real-time security policies and perform in-network certificate validation, ensuring that communication is solely allowed to trusted OPC UA devices. Positioned above the Data Plane Layer, the IOTA Clients Layer acts as a middleware connecting the DLT with the devices in both Layer \Circled{1} and \Circled{2}, facilitating seamless management of OPC UA certificate transactions. At the top of the architecture, the IOTA Layer provides a secure, immutable ledger that underpins the entire system, handling the issuance and revocation of OPC UA certificates.
\subsubsection{Device Layer}
The Device Layer consists of OPC UA clients and servers already deployed within the ICS network, as described in Section~\ref{sec:threatmodel}. Each device is configured by the network administrator to use the Sign\&Encrypt security mode, avoid deprecated security policies, and is equipped with a self-signed certificate or one issued by a trusted Certification Authority. 
We expect that OPC UA devices can perform all certificate validation steps outlined in the OPC UA standard specification~\cite{OPC10000-4}, excluding the "Trust List Check" for the reasons outlined in Section~\ref{sec:problem}. This step involves verifying whether an OPC UA certificate is trusted by determining if it is included in the device's trust list and, in the Pk-IOTA architecture, this validation is performed in-network at the Data Plane Layer, offloading it from end OPC UA devices.

\subsubsection{Data Plane Layer}
\label{sec:dataplanelayer}
The Data Plane Layer consists of P4 programmable switches inspecting network packets and enforcing trust on every connection establishment handshake between an OPC UA client and server. In particular, whenever an OPC UA Open Secure Channel request is sent by an OPC UA client, the programmable switch parses the OPC UA payload extracting the sender's certificate and the certificate thumbprint. Subsequently, it checks the certificate's trust status: if the certificate is untrusted or revoked, the programmable switch drops the packet, and the OPC UA connection is rejected directly at the data plane, preventing not trusted endpoints from establishing OPC UA Secure Channels in the first place. This process is then repeated with the following OPC UA Open Secure Channel response sent by the OPC UA server.

The \textit{Pk-IOTA P4 Parsing Pipeline} begins by handling the standard TCP/IP packet headers (Ethernet, IPv4, and TCP). Once the pipeline identifies that the payload is an OPC UA message, the parser transitions to a specialized state that extracts the OPC UA header and determines the OPC UA message type. If the OPC UA message type is \texttt{OPN} (indicating an Open Secure Channel), the switch parses the corresponding security header and proceeds to parse the sender's certificate and its thumbprint. As shown in Algorithm~\ref{algo:p4_cert}, the parser retrieves the certificate length from the OPC UA message and then performs the necessary endianness adjustments. Specifically, although the OPC UA specification encodes the certificate length as a big-endian integer, it must be converted to little-endian to be processed as bytes in the P4 language. Since the maximum size of a variable in P4 is 256 bytes, and the certificate may exceed this limit, the parser processes the certificate iteratively. The final parsing step involves extracting the certificate thumbprint.
As soon as the OPC UA Open Secure Channel request or response is completely parsed, the Data Plane Layer enforces in-network, real-time trust verification of OPC UA certificates within the Ingress processing pipeline. The P4 switch checks the certificate thumbprint against a dedicated match-action table (\texttt{thumbprint\_table}). This table is populated by the P4 controller, which installs rules mapping valid thumbprints to an "allow" action. If a matching thumbprint rule is found, the packet is classified as trusted, permitting the OPC UA Secure Channel establishment to continue. Conversely, if no corresponding rule exists, the data plane immediately drops the packet, thus preventing any connections with unknown or unauthorized OPC UA certificates. This mechanism effectively offloads trust-list management from individual OPC UA devices, enforcing real-time, centralized policy control at the data plane level.



\begin{algorithm} \footnotesize
\caption{\small OPC UA Certificate parsing in P4.}\label{algo:p4_cert}
\begin{algorithmic}[1]
\algrenewcommand\algorithmicrequire{\textbf{Input:}}
\algrenewcommand\algorithmicensure{\textbf{Output:}}
\algnewcommand{\algorithmicand}{\textbf{ and }}
\algnewcommand{\AND}{\algorithmicand}

\algnewcommand{\algorithmicstate}{\textbf{state }}
\algnewcommand{\algorithmictrans}{\textbf{transition select}}
\Require packet p
\Ensure opcua\_cert

\Procedure{extractCert()}{}
\State bit<2048>[100] opcua\_cert;
\State bit<32> cert\_bytes;
\State bit<32> remaining;

\State \algorithmicstate parse\_certLength
    \State\hspace{\algorithmicindent} p.extract(hdr.opcua\_certLength);
    \State\hspace{\algorithmicindent} bit<8> hex1 = hdr.opcua\_certLength[31:24];
    \State\hspace{\algorithmicindent} bit<8> hex2 = hdr.opcua\_certLength[23:16];
    \State\hspace{\algorithmicindent} bit<8> hex3 = hdr.opcua\_certLength[15:8];
    \State\hspace{\algorithmicindent} bit<8> hex4 = hdr.opcua\_certLength[7:0];
    \State\hspace{\algorithmicindent} cert\_bytes = (bit<32>)(hex4 ++ hex3 ++ hex2 ++ hex1);
    \State\hspace{\algorithmicindent} remaining = cert\_bytes;
    \State\hspace{\algorithmicindent} \algorithmictrans(cert\_bytes)
        \State\hspace{\algorithmicindent}\hspace{\algorithmicindent} 0 : drop;
        \State\hspace{\algorithmicindent}\hspace{\algorithmicindent} \_ : check\_cert\_length;

\State \algorithmicstate check\_cert\_length
    \State\hspace{\algorithmicindent} \algorithmictrans(cert\_bytes > 255)
        \State\hspace{\algorithmicindent}\hspace{\algorithmicindent} false : parse\_certificate\_ending\_part\_only;
        \State\hspace{\algorithmicindent}\hspace{\algorithmicindent} true : parse\_certificate;

\State \algorithmicstate parse\_certificate
    \State\hspace{\algorithmicindent} packet.extract(opcua\_cert.next);
    \State\hspace{\algorithmicindent} remaining = remaining - 256;
    \State\hspace{\algorithmicindent} \algorithmictrans(remaining > 255)
        \State\hspace{\algorithmicindent}\hspace{\algorithmicindent} false : parse\_certificate\_ending\_part;
        \State\hspace{\algorithmicindent}\hspace{\algorithmicindent} true : parse\_certificate;

\State \algorithmicstate parse\_certificate\_ending\_part
    \State\hspace{\algorithmicindent} packet.extract(opcua\_cert, (bit<32>)(remaining * 8));

\State \algorithmicstate parse\_certificate\_ending\_part\_only
    \State\hspace{\algorithmicindent} packet.extract(opcua\_cert, cert\_bytes);
\State \Return opcua\_cert
\EndProcedure
\end{algorithmic}
\end{algorithm}

\subsubsection{Iota Clients Layer}
To complement the real-time certificate enforcement performed by the Data Plane Layer, the IOTA Clients Layer provides an interface for OPC UA network administrators to manage and distribute OPC UA trusted certificates via the IOTA Tangle. In this architecture, the P4 controller itself acts as a listener to relevant IOTA transactions originating from the administrator wallet. Whenever these transactions carry an OPC UA certificate for issuance or revocation, the controller updates the \texttt{thumbprint\_table} entries in the P4-programmable switch accordingly by either installing a match-action rule to allow OPC UA connections for a newly trusted certificate or removing a rule to prevent future connections for a revoked one. This mechanism ensures that trust relationships are tamper-proof and maintained in near-real time throughout the ICS network, offloading individual OPC UA devices from the complexities of certificate distribution.
From the Stardust release onward, IOTA enabled the implementation of smart contracts by allowing interoperability between two layers: Layer 1 (L1), based on IOTA's Tangle, and Layer 2 (L2), which works as a sidechain infrastructure and supports Ethereum virtual machines (EVMs)~\cite{iotasmartcontr2021}.

This dual-layer architecture enables different approaches to managing certificates, with each layer providing distinct advantages in terms of speed, computational requirements, and functional capabilities. We propose the Pk-IOTA architecture with two different interactions with the IOTA layer: MQTT interactions with the Tangle (L1) and Smart contract (L2); the two solutions are distinct, and it can be chosen which of the two to use, as they reach the same objectives but with different requirements and effects, both in terms of performance and functionalities.

In the L1 implementation, IOTA clients utilize the \texttt{iota-sdk} MQTT Python library to facilitate the exchange of transactions between devices. This framework allows the sender to read certificates sequentially and transmit them as the payload of a Tangle transaction. Receivers, meanwhile, listen for transactions tagged with specific labels, such as "certificate", using the MQTT protocol.
The certificate data can be encoded within the transaction's payload in various formats, together with a boolean flag indicating whether the transaction involves the issuance of a new certificate or its revocation. To ensure the integrity of the process and filter out unauthorized submissions, receivers only consider transactions originating from a designated administrator: clients verify the sender's signature by checking that the associated public key is among the pre-authorized keys (i.e., one of the administrator's keys). Each client implements both the sender and listener functionalities to enable bidirectional communication. This implementation leverages the inherent scalability and efficiency of the Tangle, allowing us to assess baseline performance in terms of computational overhead.\\
In the L2 implementation, clients use Node.js along with the \texttt{web3} library to interact with SCs deployed on the L2 sidechain. These SCs provide various functionalities to automate certificate management. They maintain a list of all valid certificates, automatically removing entries when certificates expire or are revoked.
Certificates are managed by specific functions: \texttt{addCertificate(cert, expireDate)} and \texttt{revokeCertificate(cert)}. These functions update the local list of certificates and their statuses. Moreover, when changes occur in the list, the SC emits an event to notify all IOTA clients, ensuring real-time updates.
The SC maintains a tamper-proof and valid list of certificates and facilitates certificate management and automation, such as during client initialization. It offers a function, \texttt{getAllCertificates()}, that retrieves all valid certificates and sends them to clients. As a result, clients only need to use the Node.js frontend application to interact with the SC and receive event notifications. Although this approach introduces a higher computational load due to the additional complexity of smart contract execution, it offers greater functionality and automation.

\subsubsection{Iota Layer}
The Iota Layer contains the IOTA Tangle, which is employed to provide tamper-proof and immutable records of certificate information, including revocations. Clients can interact with the Tangle either by running a local IOTA node or by connecting to publicly available nodes provided by IOTA as a service. IOTA is equipped with load balancing to optimize performance and ensure reliable access to their publicly available nodes.\\
Operating a local IOTA node offers direct, real-time visibility into the Tangle and no fees for transactions containing OPC UA certificates, enhancing control and reliability. However, this setup requires substantial computing resources, which may be challenging for devices with limited processing power or storage capacity. Therefore, clients must carefully weigh the benefits of direct monitoring against the resource demands when choosing between a local node and a public one.
Local nodes utilize the Hornet daemon, managing L1 network assets and related tools. Clients can connect to both public testnets and mainnets in the L1 network. For SC capabilities, the Wasp daemon is used alongside Hornet to handle L2 assets. However, due to current security policies set by the IOTA community, running a local L2 node connected to the official IOTA public L2 sidechain is not possible. Consequently, clients requiring SC functionality can only connect to public L2 nodes provided by IOTA via their load balancing service or deploy a private L2 sidechain running local nodes. The choice depends on the specific trade-offs between resource availability, the desired control over the blockchain environment, and eventual fees.

\subsection{Workflow}
\label{subsec:workflow}

\begin{figure*}
    \centering
    \includegraphics[width=1.0\linewidth]{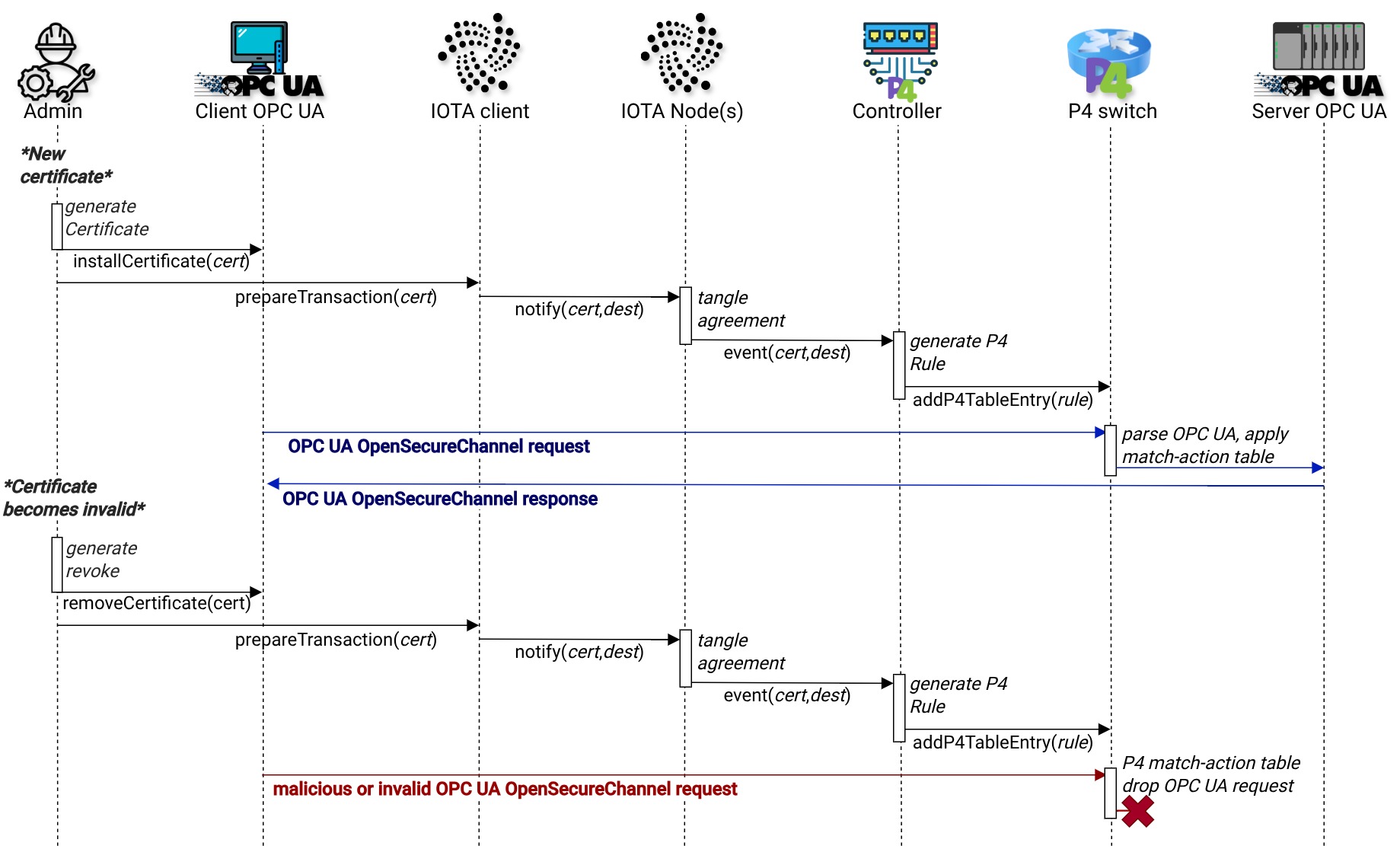}
    \caption{Workflow of the PK-IOTA certificate management process for adding and revoking certificates.}
    \label{fig:workflow}
\end{figure*}

Figure~\ref{fig:workflow} illustrates how new OPC UA devices are added as trusted entities in the OPC UA network and how OPC UA certificates are revoked. The workflow unfolds through coordinated interactions among the network administrator, the IOTA network, the P4 controller, and the P4 switch, thus ensuring that certificate updates are securely propagated throughout the 4 layers of the Pk-IOTA architecture.

\subsubsection{Insertion of a Certificate}
Initially, the network administrator generates or obtains a valid OPC UA certificate for the new client or server. During the process of provisioning the certificate to the OPC UA device, the network administrator automatically interacts with the IOTA network using an IOTA client to register the new certificate as trusted. This certificate is published on the IOTA network through the IOTA Clients Layer, either to the Tangle on Layer1 via an MQTT-based transaction or to a smart contract on Layer2.
Once the certificate transaction is successfully processed on the IOTA network, the local IOTA client on the P4 controller is notified. In response, the controller extracts the certificate's thumbprint and installs the corresponding entry in the \texttt{thumbprint\_table} of the P4 switch via the \texttt{TableEntry} P4Runtime API~\cite{p4runtimespec}.
These rules are enforced by the P4 switch, which controls the traffic between OPC UA clients and the server, allowing the new device to establish OPC UA secure channels with other trusted entities. Any future connection request involving the new certificate is accepted immediately, provided its thumbprint matches an entry in the switch's table.

\subsubsection{Revocation of a Certificate}
Revocation follows a similar process. The administrator, upon identifying a potentially compromised or expired certificate, sends a revocation transaction through the IOTA Clients Layer. On Layer1, the Tangle processes and broadcasts this update, while on Layer2, a smart contract removes the certificate from its valid list and emits an event indicating its revocation. In both cases, the P4 controller detects the revocation message through its local IOTA client and removes the certificate's thumbprint from the \texttt{thumbprint\_table}. As a result, any subsequent attempt to initiate an OPC UA connection with that certificate is dropped at the data plane, effectively cutting off compromised devices from the network.

\section{Evaluation}
\label{sec:eval}
This Section evaluates Pk-IOTA's overall efficacy to assess whether it can be applied to real-world Industry 4.0 scenarios for securing OPC UA communications with real-time constraints. In particular, we focus on empirically answering the following evaluating questions: 
\begin{enumerate}
    \item[\textbf{Q1}] How much in-network overhead does Pk-IOTA introduce on the data plane for validating OPC UA certificates during the OPC UA security handshake?     
    \item[\textbf{Q2}] How efficiently can Pk-IOTA propagate and manage OPC UA certificates through the IOTA Tangle and how do the Layer 1 and Layer 2 approaches compare in terms of latency, resource consumption, and functionality?
    \item[\textbf{Q3}] What is the total resource consumption of the various components of the Pk-IOTA architecture, and how does this impact runtime performance under realistic operational conditions?
\end{enumerate}
We begin by outlining the experimental setup used to conduct the tests needed to answer these questions.

\subsection{Experimental Setup}

To evaluate the proposed Pk-IOTA architecture, we set up a physical testbed. It adheres to the Purdue Enterprise Reference Architecture, maintaining the conventional separation between the Information Technology and Operational Technology domains. 

The IT domain is implemented using a single high-performance workstation, representing the Enterprise Zone of the testbed. This workstation, a Dell PC equipped with an Intel i7 processor and 64 GB of RAM, hosts the OPC UA client (implemented with opcua-asyncio\footnote{\url{https://github.com/FreeOpcUa/opcua-asyncio}}) and the IOTA client for managing transactions. It represents the HMI and main workstation for the system administrator.

The interconnection between the IT and OT domains is realized through an Aruba Switch. The OT segment is structured into three layers, omitting the SCADA Level 2 of the Purdue Reference Architecture due to the small scale of the testbed.
The first layer corresponds to Level 0 in the Purdue model and includes sensors and actuators\footnote{\url{https://www.dexterindustries.com/grovepi/}}. Additionally, we employ an embedded IoT device capable of using different industrial protocols\footnote{\url{https://sklep.inveo.com.pl/en/monitoring/49-nano-temp.html}}. The second layer, corresponding to Level 1, is composed of a Revolution Pi (RevPi) device\footnote{\url{https://revolutionpi.com/en/revolution-pi-series}}, an open-source industrial-grade computer used to emulate various components of an industrial network such as PLCs or Edge Nodes. The RevPi was chosen for its versatility in simulating industrial protocols like OPC UA and Modbus, as well as for its compatibility with P4 bmv2 virtual switches.
The third layer, corresponding to Level 3 of the Purdue Architecture, is represented by an edge node equipped with an Intel i7 processor and 32 GB of RAM. This edge node hosts the IOTA client, which observes certificate transactions, and a custom P4 controller presented in our previous work~\cite{al2024unleashing,rinieri2024network}. The edge node is connected to the Aruba Switch, enabling its interaction with the IOTA blockchain.

\subsection{In-Network certificate validation Overhead (Q1)}
To quantify the overhead introduced by in-network certificate validation within Pk-IOTA, we conducted a series of tests to measure packet processing and packet dequeuing times. These tests analyzed the impact of P4 data plane in-network certificate validation on the OPC UA connection establishment process. Specifically, we tagged Open Secure Channel request and response packets (i.e., with \texttt{OPN} message type) within the parser pipeline with a boolean variable in the user-defined P4 metadata structure. Then, in the egress pipeline, we measured the packet processing and dequeuing times exclusively for tagged packets. The results of these computations were stored in registers, enabling a precise packet-level analysis.

\begin{figure*}
    \centering
    \includegraphics[width=0.55\textwidth]{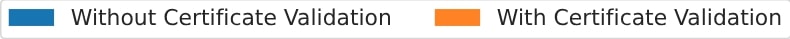}
    
    \begin{subfigure}{0.3\textwidth}
        \centering
        \includegraphics[width=\textwidth]{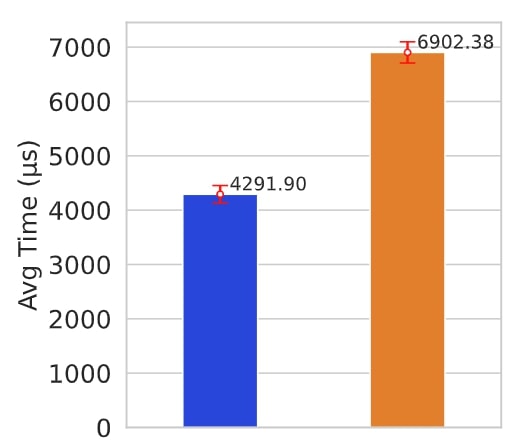}
        \caption{Average Packet Processing Time.}
        \label{fig:PPT}
    \end{subfigure}
    \hfill
    \begin{subfigure}{0.3\textwidth}
        \centering
        \includegraphics[width=\textwidth]{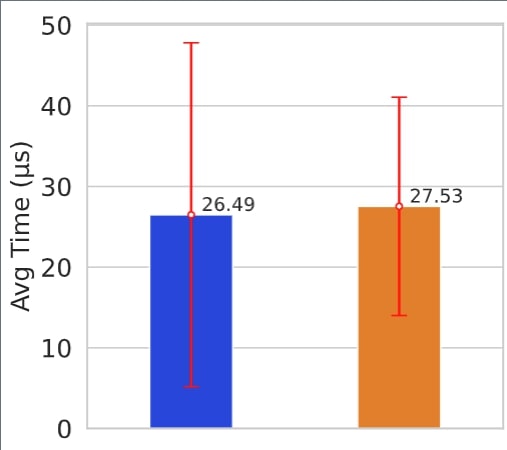}
        \caption{Average Packet Dequeuing Time.}
        \label{fig:deq}
    \end{subfigure}
    \hfill
    \begin{subfigure}{0.3\textwidth}
        \centering
        \includegraphics[width=\textwidth]{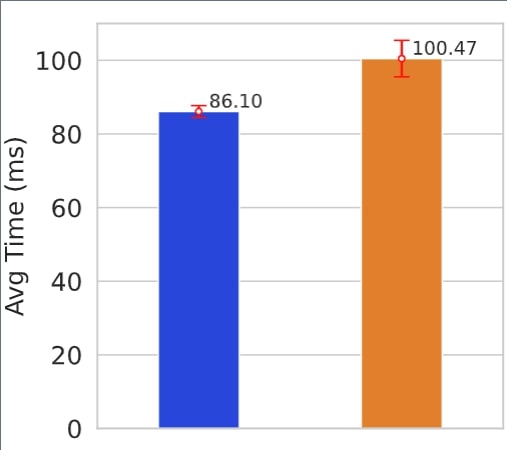}
        \caption{Average OPC UA Handshake Time.}
        \label{fig:conn}
    \end{subfigure}
    
    \caption{Comparison of average packet processing, packet dequeuing, and end-to-end handshake times for 1000 OPC UA connection attempts with and without in-network certificate validation. Packet processing and dequeuing times are computed in the P4 data plane installed on the RevPi. 
    }
    \label{fig:time}
\end{figure*}

Figure~\ref{fig:PPT} presents the packet processing times for 1000 OPC UA connection attempts, covering 2000 Open Secure Channel requests and responses. When comparing standard OPC UA traffic to those with in-network certificate validation, we observe an average increase of approximately 38\% (2600~$\mu s$) per packet. This extra processing is attributed to the additional steps required to parse and validate OPC UA certificates in the data plane as described in Section~\ref{sec:dataplanelayer}.
Figure~\ref{fig:deq} displays the packet dequeuing times for the same experimental setup. The introduction of in-network certificate validation adds only about 1.04 $\mu s$ per packet, highlighting its minimal impact on overall network throughput.
Finally, Figure~\ref{fig:conn} compares the end-to-end handshake times across 1000 OPC UA connection attempts, with and without certificate validation. The total time for performing the OPC UA security handshake increased by approximately 14\% (14 ms) on average, due to the in-network validation process. Despite this overhead, we claim that the handshake process remains within acceptable limits for real-time industrial applications.
Overall, the results confirm that the inclusion of in-network certificate validation introduces a measurable but minimal overhead~\cite{cavalieri2013analysis,kohnhauser2022feasibility}. The implementation ensures that the security of OPC UA communications is enhanced without significantly compromising the performance or real-time requirements of industrial control systems.

\subsection{OPC UA propagation delay over the IOTA Tangle (Q2)}

Pk-IOTA is designed to provide a scalable, globally distributed system for certificate management. To evaluate its feasibility and performance, we tested certificate propagation delays across both Layer 1 and Layer 2 implementations of the IOTA Tangle. In doing so, we aim to identify the trade-offs between the two in terms of delay, computational efficiency, and additional functionalities.
We deployed distinct Hornet and Wasp IOTA nodes over VMs located in different physical positions. In particular, we placed the VMs respectively over long distances (United States - Australia, Europe - Australia) and short distances (North Europe - Central Europe) to evaluate the impact of the use of this technology over networks of different sizes, that is a requirement of our distributed IIoT use-case scenario. Moreover, we performed the tests with certificates in different formats to also assess the difference in delays caused by payload size. For both L1 and L2, we performed 300 tests for each distance and each certificate format.

\begin{figure*}
    \centering
    \begin{subfigure}{0.49\textwidth}
        \centering
        \includegraphics[width=\textwidth]{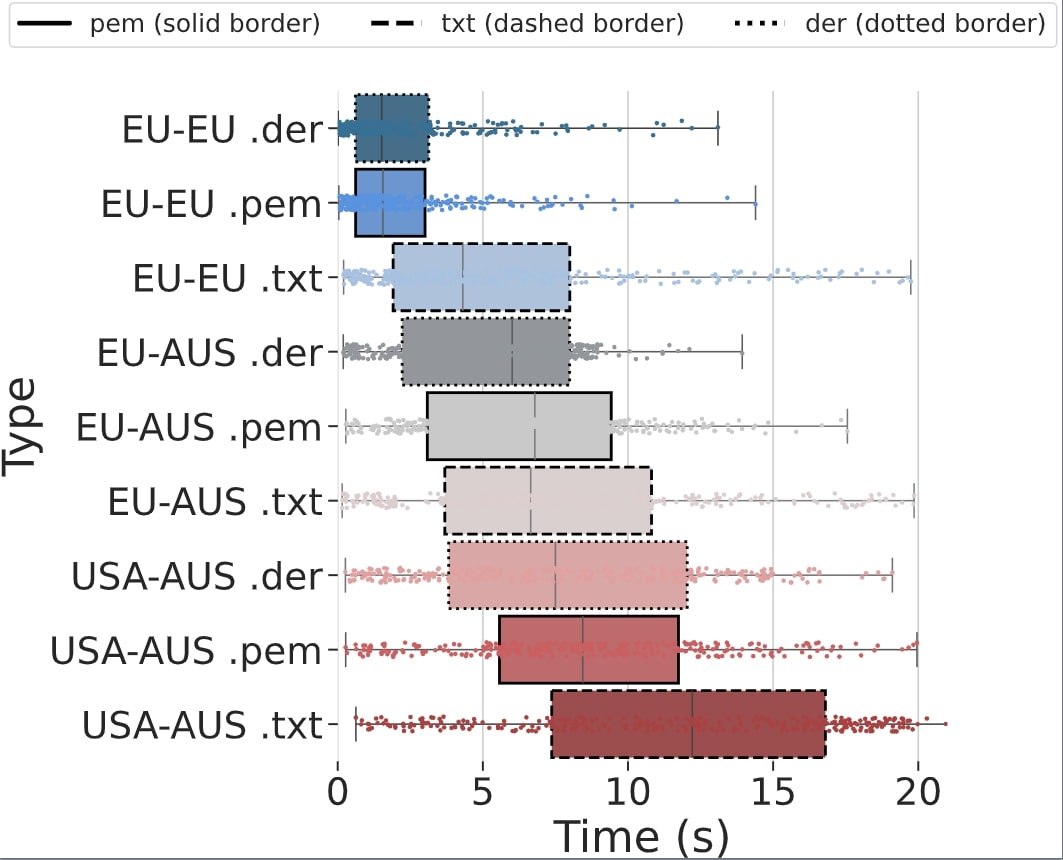}
        \caption{Time is taken to send and receive transactions in the IOTA L1 layer.}
        \label{fig:mqtt}
    \end{subfigure}
    \begin{subfigure}{0.47\textwidth}
        \centering
        \includegraphics[width=\textwidth]{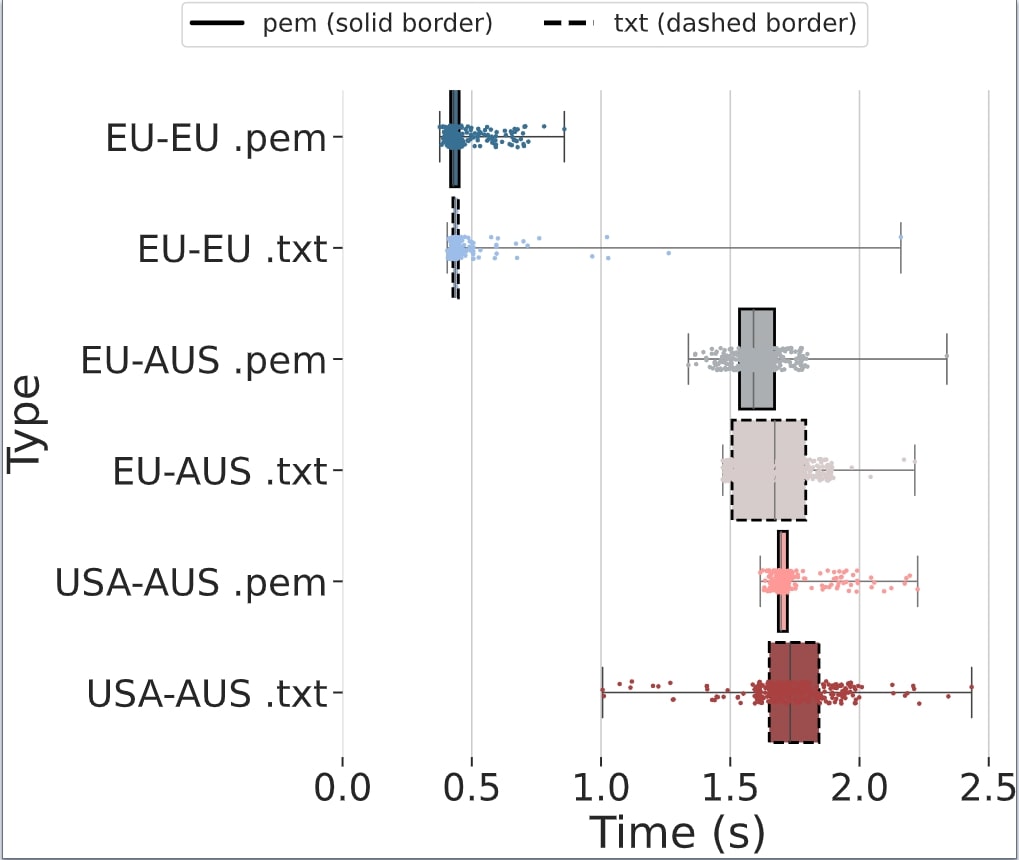}
        \caption{Time is taken to invoke the SC in the IOTA L2 layer.}
        \label{fig:sc}
    \end{subfigure}
    \caption{L1 and L2 performance comparison. The different certificate formats are distinguished by varying box border styles, while the color shades indicate the different physical distances.}
\end{figure*}

The L1 tests are performed using the Python API of the IOTA-sdk mqtt library. Transactions are sent with JSON-RPC calls to the local Hornet node of the VMs, which operates over the IOTA Testnet. The delay is considered between the local time of the sender's call to the function \texttt{send()} of IOTA-sdk and the local time at the reception of the transaction from the listener, with a previous synchronization of the two clocks.
Results of L1 tests are shown in Figure \ref{fig:mqtt}: the average transaction time is proportional to the distance between network nodes. This behavior is likely influenced not only by network delay but also by the consensus mechanism of IOTA, given the significant differences observed across cases. Furthermore, in smaller networks, the number of outliers is notably lower, reinforcing that the efficiency of the transaction confirmation mechanism depends on node proximity~\cite{fan2021performance,zander2019dagsim}. The process by which a transaction is added in IOTA primarily relies on the closest nodes, making this outcome expected, and delays caused by more distant transactions are likely also due to transaction reattachments within the DAG~\cite{conti2022survey}. However, results show that even in cases where reattachment occurs, the delay remains within 20 seconds, which is deemed acceptable since the certificate distribution is an offline process.

The L2 tests, based on SC interactions, are performed with \textit{Nodejs} projects and the \texttt{web3} library. We wrote the SC in Solidity and deployed it over a private sidechain, where all the Wasp nodes of the VMs of our tests take part. 
The delays are measured from the moment the sender calls the \texttt{sendCertificate()} function and the reception of the event by the listener. Tests are performed for the insertion case, as the revoke mechanism has an identical structure and, therefore, would give the same results in terms of delay. 
The results on Layer 2, shown in Figure \ref{fig:sc}, differ significantly from those observed on Layer 1. We found that L2 achieved approximately 75\% lower delays than L1, ranging between 1.5 and 2 seconds for long-distance cases and around 0.5 seconds for shorter distances, with no evident differences related to payload size. The low variance is likely due to the smaller private network of the sidechain and the distinct L2 consensus mechanism, where the execution of the SC relies on a sidechain that relies on a DAG, rather than attaching a transaction directly to a DAG~\cite{iotasmartcontr2021}. Since updating the chain state on the L1 DAG is decoupled from the smart contract's execution, it does not affect the smart contract's response time, allowing for quicker processing and additional functionalities, such as the \textit{getAllCertificates()} function, which retrieves all valid certificates at a given moment.

\subsection{Overall Resource Consumption (Q3)}
\begin{table*}
\centering
\caption{Comparison of memory usage, CPU usage, and power consumption for each component.}
\resizebox{0.8\textwidth}{!}{
\begin{threeparttable}
\begin{tabular}{l l rr rr rr rr}
\hline
Component & Deployed in & \multicolumn{2}{c}{Memory (MB)} & \multicolumn{2}{c}{CPU (\%)} & \multicolumn{2}{c}{Power Cons. (W)}\\
 \cmidrule(lr){3-4}\cmidrule(lr){5-6}\cmidrule(lr){7-8}
& & Mean & Max & Mean & Max & Mean & Max \\ \hline
IOTA sender & Windows Workstation & 148.17 & 154.00 & 0.61 & 0.63 & 19.18 & 19.37 \\ \hline
IOTA receiver & Edge Node & 132.22 & 170.02 & 0.23 & 0.67 & 18.53 & 19.40\\ \hline
IOTA node & Azure Cloud VM & 444.27 & 455.31 & 3.38 & 3.80 & 25.33 & 26.21 \\ \hline
OPC UA Client & Windows Workstation & 53.97 & 54.00 & 1.54 & 2.28 & 18.73 & 20.45 \\ \hline
OPC UA Server & RevPi Connect 4 & 110.85 & 112.12 & 10.31 & 11.30 & 2.15$^*$ & 2.15$^*$ \\ \hline
P4 bmv2 & RevPi Connect 4 & 233.45 & 262.76 & 29.09 & 35.75 & 2.15$^*$ & 2.15$^*$ \\ \hline
P4 Controller & Edge Node & 288.05 & 288.05 & 6.63 & 6.97 & 45.77 & 57.80 \\ \hline
\end{tabular}
\begin{tablenotes}
\tiny 
\item $^*$Considering that power consumption measurements over a RevPi require specific hardware, we estimated it using Ohm's law $P = V \cdot I$, where the voltage is obtained via \textit{vcgencmd measure\_volts} and the electric current value is 2.5A, provided to the RevPi by the \href{https://revolutionpi.com/shop/en/netzteil-meanwell-hutschiene-mdr-60-24}{DIN Rail Power Supply Unit MDR-60-24}.
\end{tablenotes}
\end{threeparttable}
}
\label{tab:resourceconsumption}
\end{table*}

Table~\ref{tab:resourceconsumption} summarizes the memory, CPU, and power consumption of each Pk-IOTA component, revealing how resource demands vary according to each component's role within the architecture. The IOTA node exhibits the highest memory footprint at 444.27 MB on average, reflecting the overhead of maintaining a direct connection to the Tangle and processing distributed certificate transactions. By contrast, the IOTA sender and receiver processes require significantly less memory (148.17 MB and 132.22 MB on average, respectively) since they only handle lightweight, client-side interactions with the ledger.
The RevPi, which functions as both an OPC UA server and a host for the bmv2 P4 switch, averages 233.45 MB of memory consumption, underscoring the computational demands of simulating industrial devices and running in-network certificate checks. This dual role also contributes to its higher CPU utilization, which averages 29.09\%. 
In terms of power consumption, the operating system of the edge node that hosts the P4 controller and IOTA client reports the highest average draw at 45.77 W. This usage reflects its dual responsibilities of continuously listening for certificate transactions and dynamically modifying match-action table rules via the P4Runtime API. Meanwhile, the RevPi's power usage, estimated through Ohm's law, remains relatively modest at 2.15 W, even though it manages industrial traffic emulation and P4 switching concurrently.
Overall, these findings confirm that while resource usage differs across components, the Pk-IOTA architecture operates within feasible limits for typical industrial applications. The results highlight that both DLT-based certificate management and in-network certificate enforcement can be supported in real time without imposing excessive strain on the underlying hardware.

\section{Discussion \& Limitations} 
\label{sec:discussion}
\subsection{Security Analysis}
The security of the OPC UA protocol has been extensively analyzed and formally proven by \cite{puys2016formal,dreier2017formally,dreier2019formally}. Building on these formal guarantees, our Pk-IOTA architecture augments OPC UA with a decentralized certificate manager and in-network certificate validation to ensure robust protection against Rogue Server, Rogue Client, and Middleperson attacks. To formally prove this, based on our threat model described in Section~\ref{sec:threatmodel}, we model the DLT as an authentic channel. By doing so, we can assume that messages published on, or retrieved from, the IOTA ledger cannot be tampered with or falsified by the attacker. While an adversary can read these messages, they cannot alter the sender's identity, that is the network administrator in our case, or the content of any transaction. This property ensures that a certificate added to the IOTA ledger for issuance or revocation arrives unmodified at all legitimate parties.
Additionally, we further model the communication link between the P4 controller and the P4 programmable switch as a secure channel, which is both confidential and authentic. An adversary cannot learn or modify any messages exchanged over this link. This prevents attackers from manipulating the content of P4 rules or forging new entries in the \texttt{thumbprint\_table}. We now proceed to illustrate how the Pk-IOTA architecture thwarts each attack.

\subsubsection{Security against Rogue Server Attacks}
Let $C_s$ be the certificate of an OPC UA server $S$, and let $T$ denote the trusted certificate set stored on the IOTA ledger. In Pk-IOTA, the P4 data plane inspects each \texttt{OpenSecureChannel} request and checks whether $C_s \in T$. If $C_s \notin T$, the request is dropped immediately, preventing the connection from being established.
Under our threat model, the Dolev--Yao attacker cannot modify the IOTA ledger or falsify transactions because we treat the ledger as an \emph{authentic channel}: the attacker can read messages but cannot alter them or spoof their origin. Similarly, the adversary cannot tamper with the secure channel between the P4 controller and the programmable switch, so it cannot install malicious rules or bypass the ledger-based trust checks. The only remaining avenues for a Rogue Server attack would be to (1) steal the ledger wallet credentials of the network administrator and add a fake certificate to $T$, or (2) compromise an existing OPC UA server directly. Both scenarios exceed the adversarial capabilities defined under our assumed Dolev--Yao attacker model. Additionally, the first scenario will involve social engineering techniques and is outside the scope of this work. The second scenario, instead, violates our assumption that the attacker does not have access to devices below Level~2 of the Purdue architecture, which is a typical location for OPC UA servers in industrial control systems. Therefore, given these channel assumptions and the ICS network segmentation, the attacker cannot succeed in creating a Rogue Server. 

\subsubsection{Security against Rogue Client Attacks}
Let $C_c$ be the certificate of an OPC UA client $C$, and let $T$ denote the trusted certificate set on the IOTA ledger. In Pk-IOTA, the P4 data plane inspects every \texttt{OpenSecureChannel} request originating from a client. If $C_c \notin T$, the packet is dropped immediately, preventing any connection attempt from proceeding.

Under our threat model, a Dolev--Yao attacker cannot alter the ledger since it is modeled as an \emph{authentic channel}; the attacker can read but cannot forge or modify transactions. Likewise, the secure link between the P4 controller and the switch is beyond the attacker's reach, so installing malicious entries in the \texttt{thumbprint\_table} is not possible. The only viable option for impersonating a legitimate client would be to (1) steal the ledger account belonging to the network administrator and add the attacker's certificate to $T$, or (2) compromise an existing OPC UA client. As for the Rogue Server attack, both scenarios lie outside our threat model, which assumes the attacker to have Dolev-Yao capabilities.

\subsubsection{Security against Middleperson Attacks}
For a Middleperson attack to succeed, the attacker must impersonate both the server $S$ (with certificate $C_s$) and the client $C$ (with certificate $C_c$). However, as shown in our analyses of Rogue Server and Rogue Client attacks, neither impersonation is feasible within the Pk-IOTA architecture. Consequently, the conditions necessary for a Middleperson attack cannot be fulfilled.

\subsection{Scalability}
The scalability of the Pk-IOTA architecture requires consideration of both the P4 programmable data plane and the IOTA blockchain. In the data plane, OPC UA certificates are parsed in 2048-bit (256-byte) chunks by means of a P4 header stack. For our experiments, we set the header stack size to 100, allowing for a maximum certificate length of $256 \cdot 100 = 25600$ bytes. Since the header stack size is defined by a 32-bit unsigned integer in the P4 language specification, this limit can be significantly increased. This is a parameter whose value should be taken into consideration at deployment time to balance resource usage and performance.

We further configured the \texttt{thumbprint\_table} in our P4 pipeline to hold up to 1024 match-action table entries. Each entry corresponds to the certificate thumbprint of an OPC UA device. We expect this capacity to suffice for most industrial networks, yet it can be expanded based on the capabilities of the P4 programmable switch. The RevPi we employed, for instance, has 32 GB of available memory, and related studies show that modern programmable switches typically provide abundant memory for match-action table entries~\cite{chen2020implementing}.

With respect to IOTA, instead, our solution behaves differently depending on whether it operates on the L1 or the L2 of the IOTA network. 
On L1, scalability is one of Tangle's core strengths. The IOTA Tangle is designed to improve in performance as the network grows, meaning that as more certificates are managed and more nodes participate, the throughput of the system scales effectively~\cite{zander2019dagsim}. Therefore, even with a significant increase in the number of certificates, the solution remains sustainable in terms of performance and cost. However, it is worth noting that operations on L1 may occasionally face network-wide confirmation delays in periods of extremely high activity, as noticed during our performance tests.
On L2, scalability depends on the architecture of the private sidechain or the chosen public L2 solution, as the chain settings are configurable at deployment time, allowing for tailored adjustments. When using a public L2 sidechain, scalability is influenced by that platform's fees and network congestion, which can become a limitation if certificate operations increase dramatically. However, in our architecture and specific PKI use case, the number of certificates is not expected to be very high, mitigating potential scalability concerns.
\vspace{-0.25cm}

\subsection{Costs of the Pk-IOTA Infrastructure}
When discussing the theoretical monetary costs of our PKI implementation for OPC UA clients and servers on the IOTA blockchain, it is essential to distinguish between L1 and L2 operations. For L1, the Tangle feeless nature ensures that registering or revoking certificates does not incur direct transaction costs. Moreover, sending transactions on L1 does not require additional infrastructure costs unless one opts to operate a local IOTA Hornet node for greater control or independence, though this is not strictly necessary. 
However, if the implementation relies on L2, a private sidechain would need to be established to maintain feeless operations. The costs would primarily stem from maintaining the L2 infrastructure, including hosting and operational expenses. Alternatively, relying on a public L2 sidechain typically involves transaction fees. Thus, the monetary costs depend on the choice between sustaining a private L2 infrastructure or paying fees for a public L2 network. Nevertheless, from the performance tests conducted, the computational resource costs required to sustain the private IOTA infrastructure have proven negligible and are entirely manageable by the workstations operating as OPC UA clients and servers.
Finally, when choosing between implementing a private IOTA infrastructure or relying on a public one, it is crucial to consider that the latter solution could expose sensitive information about the ICS.

\section{Related Works}
\label{sec:relatedwork}
OPC UA certificate management is a widely recognized challenge in industrial communication security, leading to numerous research efforts aimed at simplifying or automating the associated processes \cite{kohnhauser2021security,kohnhauser2022secure,kohnhauser2022feasibility,meier2022secure}. For instance, the work by Atutxa et al.~\cite{ATUTXA2023103802} demonstrates the potential of in-network validation of  Datagram Transport Layer Security (DTLS) certificates to enhance both the efficiency and security of IIoT communications. This approach closely aligns with Pk-IOTA's objectives, which combine blockchain technology with P4-based in-network processing to ensure robust certificate management and validation.
Beyond the scope of OPC UA, many researchers have explored blockchain-based approaches for implementing PKIs, such as Certcoin~\cite{fromknecht2014certcoin} and its privacy-aware extension~\cite{axon2016pb}. They are deployed over different public blockchains: Bitcoin, in CertCoin~\cite{fromknecht2014certcoin}, Ethereum in LightCert4IoT~\cite{garba2023lightcert4iots}, Hyperledger Fabric in DPKI~\cite{papageorgiou2020dpki}; however, they follow the same core concept: leveraging the blockchain to distribute the certification management over different nodes but still guaranteeing the authenticity and security of the infrastructure. Toorani et al.~\cite{toorani2021decentralized} implement a dynamic PKI that removes reliance on centralized Certificate Authorities by using blockchain and consensus mechanisms for secure public key registration, revocation, and verification, whereas \cite{kubilay2019certledger} exploits the blockchain to achieve a PKI with Certificate Transparency.
Most of the works use classic blockchains as an append-only storage~\cite{wang2020blockchain,lewison2016backing,matsumoto2017ikp}, thus not considering the advantages that DAG-based blockchains could bring. Only a few works leverage a DAG-based ledger for managing certificates~\cite{tesei2018iota, wang2022dag}, but they lack smart contracts integration and are based on initial versions of IOTA, which need a central coordinator. However, they confirmed the utility and advantages of DAG-based PKI, such as no transaction cost, low energy consumption, scalability, and parallelization.
By building on newer IOTA versions, Pk-IOTA expands these concepts to include both L1 and L2 functionalities, aiming to provide greater flexibility, improved performance, and richer OPC UA certificate management features.

\section{Conclusion} 
\label{sec:conclusion}
In this paper, we proposed Pk-IOTA, a novel architecture integrating programmable data planes and blockchain technology to address the pressing challenges of certificate management in OPC UA deployments within Industry 4.0 environments. 
By joining in-network certificate validation with the IOTA-based certificate management, we ensure that trust decisions remain consistent throughout the ICS network, while end devices remain focused on core OPC UA functionalities rather than managing local certificate lists. Under our threat model, we formally demonstrate that a Dolev--Yao adversary cannot successfully perform OPC UA Rogue Server, Rogue Client, or Middleperson attacks.

Our evaluation, conducted on an ad hoc physical testbed, demonstrates that the in-network certificate validation performed by P4 programmable devices introduces minimal overhead on the OPC UA security handshake, with the average time increasing by only 14\%. Our solution does not interfere with normal operations, thereby preserving real-time performance, which is essential in ICS environments.
Using IOTA's distributed ledger, we provide an immutable and tamper-proof mechanism for certificate issuance and revocation. We demonstrated that this approach maintains a low computational overhead, ensuring efficient operations even in resource-constrained environments while effectively mitigating the SPoF inherent in centralized PKI solutions.

\section*{Acknowledgments}
\noindent This work was partially supported by project SERICS (PE00000014) under the MUR National Recovery and Resilience Plan funded by the European Union - NextGenerationEU.

\bibliographystyle{IEEEtran}

\bibliography{biblio}

\end{document}